\documentstyle[12pt]{article}

\centerline{\bf  {COUPLING NONLINEAR $\sigma$-MODELS}}
\centerline {\bf{TO RELAXED YANG-MILLS SUPERMULTIPLETS}}
\centerline {\bf{IN $(2,0)$-SUPERSPACE}}
\vspace{0.7cm}
\centerline{M. S. G\'oes-Negr\~ao$^{{\star}{\dag}}$\footnote{negrao@gft.ucp.br}, J. A. %%@
Helay\"el-Neto${^{\star\dag}}$\footnote{helayel@gft.ucp.br}}
\centerline{M. R. Negr\~ao$^{{\star}{\dag}}$\footnote{guida@gft.ucp.br}}
\vspace{.5cm}
\centerline{{\it{$^{\star}$Universidade Cat\'olica de Petr\'opolis (UCP-GFT)}}}
\centerline{{\it{ Rua Bar\~ao do Amazonas 124, 25685-070, Petr\'opolis, RJ, Brazil}}}
\vspace{0.3cm}  
\centerline{{\it{$^{\dag}$Centro Brasileiro de Pesquisas F\'{\i}sicas (CBPF-MCT)}}}
\centerline{{\it{ Rua Dr. Xavier Sigaud 150, 22290-180, Rio de Janeiro, Brazil}}}
\vspace{0.5cm}

{
\begin{document} 
 
\hspace\parindent

\begin{abstract} 
\vspace{0.3cm} 
 
{\footnotesize{\bf Following a previous work on Abelian (2,0)-gauge theories, one reassesses %%@
here the task of coupling (2,0) relaxed Yang-Mills superpotentials to a (2,0)-nonlinear %%@
$\sigma$-model, by gauging the isotropy or the isometry group of the latter. One pays special %%@
attention to the extra ``chiral-like'' component-field gauge potential that comes out from the %%@
relaxation of superspace constraints.}} 
\end{abstract}

 
\newpage

In a previous paper \cite{20SYM}, one hs investigated the dynamics and the couplings
of a pair of Abelian vector potentials of a class of $(2,0)$-gauge super
multiplets(\cite{hull}-\cite{dine}) whose symmetry lies on a single U(1) group. 
Since a number of interesting features came out, it was a natural question to ask how these %%@
fields would behave if the non-Abelian version of the theory was to be considered.

We can see that some subtle changes indeed occur. As we wish to make a full
comparison between the two aspects (Abelian and non-Abelian) of the same sort of
theory, all the general set up of the original formulation was kept. 
 
The fundamental non-Abelian matter superfields are the 
scalar and left-handed spinor superfields, both subject to the chirality constraint; their %%@
respective component-field expressions are given by: 
\begin{eqnarray} 
{\Phi}^{i}(x;\theta,\bar \theta)&=& 
e^{i\theta{\bar\theta}{\partial}_{++}}(\phi^{i} +\theta 
\lambda^{i}),\nonumber\\ 
{\Psi}^{I}(x;\theta,\bar \theta)&=& 
e^{i\theta{\bar\theta}{\partial}_{++}}(\psi^{I} +\theta \sigma^{I}), 
\label{c5} 
\end{eqnarray} 
the fields $\phi^{i}$ and $\sigma^{I}$ are scalars, whereas
$\lambda^{i}$ and $\psi^{I}$  stand respectively for
right- and left-handed Weyl spinors. The indices ${i}$ and ${I}$ label the
representations where the correspondenting matter fields are set to transform under the %%@
Yang-Mills group. 
 
We present below the gauge transformations of both $\Phi$
and $\Psi$, assuming that we are dealing with a compact and simple gauge
group, ${\cal G}$, with generators ${G_{a}}$ that fulfill the algebra
$[G_{a},G_{b}]$=$if_{abc}G_{c}$. The transfomations read as below:  
\begin{eqnarray} 
{\Phi}'^{i}=R(\Lambda)^{i}_{j}\Phi^{j}, \hspace{1cm}
{\Psi}'^{I}=S(\Lambda)^{I}_{J}\Psi^{J},  \label{c8} 
\end{eqnarray}  
where $R$ and $S$ are matrices that respectively represent a gauge  group element in the
representations under which $\Phi$ and $\Psi$  transform. Taking into account
the chiral constraints on $\Phi$ and $\Psi$,  and bearing in mind the
exponential representation for $R$ and $S$ in terms of the group generators, 
we find that the gauge parameter superfields, $\Lambda^{a}$, must satisfy the 
same sort of constraint, namely, they are chiralscalar superfields. 
 
The kinetic action for ${\Phi^{i}}$ and ${\Psi^{I}}$ can be made invariant 
under the local transformations (\ref{c8}) by minimally coupling 
gauge potential superfields, $\Gamma^{a}_{--}(x;\theta,{\bar \theta})$ 
and $V^{a}(x;\theta,{\bar \theta})$, according to the minimal coupling 
prescriptions: 
\begin{eqnarray} 
S_{inv}=\int d^2 x d\theta d{\bar \theta} \{i[{\bar \Phi}e^{hV} 
(\nabla_{--} \Phi)- ({\bar\nabla}_{--} 
\bar\Phi)e^{hV}\Phi]+{\bar\Psi}e^{hV}\Psi\}, 
\label{c10} 
\end{eqnarray} 
as it has already been done in ref. \cite{20SYM}.
 
The infinitesimal gauge transfomations for $V^{a}$ and ${\Gamma}^{a}_{--}$ are given
by \begin{eqnarray}
{\delta}V^{a} = \frac{i}{h}{({\bar\Lambda} - {\Lambda})}^{a} -
\frac{1}{2}f^{abc}({\bar\Lambda} + {\Lambda})_{b}V_{c}
\label{dv}
\end{eqnarray}
and
\begin{eqnarray}
{\delta}{\Gamma}_{--}^{a} = - f^{abc}{\Lambda}_{b}{\Gamma}_{c--} +
\frac{1}{g}{\partial}_{--}{\Lambda}^{a}.
\label{dg}
\end{eqnarray}

No derivative acts on the ${\Lambda}^{a}$'s in eq.(\ref{dv}), which suggests
the possibility of choosing a Wess-Zumino gauge for $V^{a}$. If such a
choice is adopted and if the superfield $V$ is kept the same as in ref. \cite{20SYM}, the same %%@
identifications done in the Abelian case hold but once the Wess-Zumino gauge is left behind, %%@
the superfield-strength is lost. In order to make it possible to write down a %%@
superfield-strength in any gauge chosen, it is necessary to redefine our superfield $V$ as %%@
follows:
\begin{eqnarray}
V^a(x, \theta,\bar\theta)= C^a + \theta \xi - \bar\theta \bar\xi + \theta{\bar\theta} \hat %%@
v_{++},
\end{eqnarray}
where the $\hat v_{++}$ is given by
\begin{eqnarray}
\hat v_{++}= v^{a}_{++} + \frac{ig}{2}f^{abc}{\xi}_{b}{\bar\xi}_{c}
\end{eqnarray}
is the real gauge field of the theory.
The $\theta$-expansion for $\Gamma^{a}_{--}$ is the same as in ref. \cite{20SYM} and reads:
\begin{eqnarray} 
{\Gamma}^{a}_{--}(x;\theta,{\bar 
\theta}) &=& \frac{1}{2}(A^{a}_{--} + iB^{a}_{--}) + i\theta (\rho^{a} +
i\eta^{a}) \nonumber\\ &+& i{\bar \theta}  (\chi^{a} + i\omega^{a}) +
\frac{1}{2}\theta{\bar \theta}(M^{a}+iN^{a}).  
\label{gama--}  
\end{eqnarray} 
 $A^{a}_{--}$, $B^{a}_{--}$ and $v^{a}_{++}$ are the light-cone components of
the  gauge potential fields; $\rho^{a}, \eta^{a}, \chi^{a}$ and $\omega^{a}$
are  left-handed Majorana spinors; $M^{a}, N^{a}$ and $C^{a}$ are real scalars
and  $\xi^{a}$ is a complex right-handed spinor. 

The gauge transformations for the $\theta$-component fields read as below:  
\begin{eqnarray}  
\delta C^a &=& - f^{a}_{bc}({\Re e}\alpha^{b})C^{c}, \nonumber\\  
\delta \xi^a &=& -\frac{i}{g}\beta - f^{a}_{bc}({\Re e}\alpha^{b}) \xi^{c} %%@
-\frac{1}{2}f^{a}_{bc}\beta^{b}C^{c},\nonumber\\  
\delta v^{a}_{++}&=&\frac{2}{g} {\partial}_{++} {\Re e}\alpha^{a} - f^{a}_{bc}
({\Re e}\alpha^{b})v^{c}_{++}, \nonumber\\  
\delta A^{a}_{--}&=&\frac{2}{g}{\partial}_{--} ({\Re e}\alpha^{a}) - f^{a}_{bc}({\Re e} %%@
\alpha^{b})A^{c}_{--},  \nonumber\\ 
\delta B^{a}_{--}&=& - f^{a}_{bc}({\Re e}\alpha^{b})B^{c}_{--}, \nonumber\\  
\delta \eta^{a} &=& - f^{a}_{bc}({\Re e} \alpha^{b})\eta^{c}+\frac{1}{2}f^{a}_{bc}({\Re e} %%@
\beta^{b})A^{c}_{--}\nonumber\\&-&\frac{1}{2}f^{a}_{bc}({\Im m} \beta^{b})B^{c}_{--} %%@
-\frac{1}{g}\partial_{--}{\Re e}\beta^{a}, \nonumber\\  
\delta\rho^{a} &=& - f^{a}_{bc}({\Re e}\alpha^{b})\rho^{c}-\frac{1}{2}f^{a}_{bc}({\Re e} %%@
\beta^{b})B^{c}_{--}, \nonumber\\  
\delta M^{a} &=& - f^{a}_{bc}({\Re e}\alpha^{b})M^{c} + f^{a}_{bc}({\partial_{++}({\Re e} %%@
\alpha}^{b}) B^{c}_{--} -2f^{a}_{bc}({\Re e}\beta^{b})\omega^{c}\nonumber\\
\delta N^{a} &=& \frac{2}{g} {\partial}_{++}{\partial}_{--} {\Re e}\alpha^{a} -
f^{a}_{bc}({\Re e}\alpha^{b})N^{c} - f^{a}_{bc}({\partial_{++}{\Re e}\alpha}^{b})A^{c}_{--} %%@
\nonumber\\&+&  2f^{a}_{bc}({\Re e}\beta^{b})\chi^{c}- 2 f^{a}_{bc}({\Re %%@
e}\beta^{b})\omega^{c} -2f^{a}_{bc}({\Im m}\beta^{b})\omega^{c} \nonumber\\
\delta \chi^{a} &=& -f^{a}_{bc}\alpha^{b}\chi^{c},\nonumber\\ 
\delta \omega^{a} &=& -f^{a}_{bc}\alpha^{b}\omega^{c} 
\label{t}  
\end{eqnarray} 
The gauge variations suggest that the $v^{a}_{++}$-component 
could be identified as the light-cone partner of $A^{a}_{--}$, 
\begin{eqnarray} 
v^{a}_{++} 
\equiv A^{a}_{++}. 
\label{ve++} 
\end{eqnarray} 
This procedure yields two component-field gauge potentials: $A^{\mu} 
\equiv (A^{0}, A^{1})=(A^{++};A^{--})$ and $B_{--}$; the latter without the $B_{++}$ partner %%@
just as it happened in the Abelian case. 
 
To discuss the field-strength superfields, we start analysing the algebra of
the gauge covariant derivatives. 

 After doing so we find out that the gauge field, $A_{\mu}$, has its field
strength, $F_{\mu\nu}$, located at the ${\theta}$-component of the combination
${\Omega}{\equiv} W_{-}+{\bar U}_{-}$, where 
\begin{eqnarray}
{[{\nabla}_{+},{\nabla}_{--}]} &\equiv& W_{+} = -igD_{+}{\Gamma}_{--} -
{\partial}_{++}{\Gamma}_{+} -ig[{\Gamma}_{+},{\Gamma}_{--}], %%@
\nonumber\\{[{\bar\nabla}_{+},{\nabla}_{--}]} &\equiv& U_{-} = -ig{\bar
D}_{+}{\Gamma}_{--}.
\end{eqnarray}

This suggests the following kinetic action for the Yang-Mills sector: 
\begin{eqnarray}
S_{YM} &=&
\frac{1}{8g^{2}}{\int}d^{2}xd{\theta}d{\bar\theta}Tr{\Omega{\bar\Omega}}
\nonumber\\
&=& {\int}d^{2}xTr[\frac{-1}{4}F_{\mu\nu}F^{\mu\nu} +
\frac{i}{8}{\Sigma}\stackrel{\mathrm{\leftrightarrow}}{\partial}_{++}{\bar\Sigma} +
\frac{1}{8}M^{2}\nonumber\\
&+& \frac{1}{8} (\partial_{++} \partial_{--}C)^{2}+\frac{1}{4}(\partial_{++}\partial_{--}C)M]+ %%@
interactions,  
\end{eqnarray}  
where ${\Sigma={\rho +i\eta} + {\bar\chi} -i{\bar\omega}}$ and
$A\stackrel{\mathrm{\leftrightarrow}}{{\partial}}B\equiv(\partial 
A)B-A(\partial B)$.

Choosing now a supersymmetry-covariant gauge-fixing, ins tead of the
Wess-Zumino, we propose the following gauge-fixing term in superspace:
\begin{eqnarray} 
S_{gf} = -\frac{1}{2\alpha}{\int}d^{2}xd{\theta}d{\bar\theta}Tr[{\Pi}{\bar\Pi}]
\end{eqnarray}
where ${\Pi}=-iD_{+}{\Gamma}_{--} + \frac{1}{2}D_{+}{\partial}_{--}V$. With this, the %%@
gauge-fixing Lagrangian became 
\begin{eqnarray}
S_{gf} = &=& -\frac{1}{2\alpha}{\int}d^{2}x\{[({\partial}_{\mu}A^{\mu})^{2} +
({\partial}_{\mu}A^{\mu})N + \frac{1}{4}N^{2}] \nonumber\\
&+& \frac{1}{4}[M^{2} -
2M({\partial}_{++}B_{--}) + ({\partial}_{++}B_{--})^{2} %%@
+(\partial_{++}\partial_{--}C)^{2}\nonumber\\
&+& 2M(\partial_{++}\partial_{--}C)(\partial_{++}B_{--})-2M(\partial_{++}\partial_{--}C)] %%@
\nonumber\\ 
&-&i({\rho + i\eta})\stackrel{\mathrm{\leftrightarrow}}{\partial}_{++}({\bar\rho
-i\bar \eta})\} + interactions. 
\label{gf}
\end{eqnarray}

So, the total action reads as follows:
\begin{eqnarray}
S &=& {\int}d^{2}xTr\{-\frac{1}{4}F_{\mu\nu}F^{\mu\nu}
-\frac{1}{2\alpha}({\partial}_{\mu}A^{\mu})^{2} -\frac{1}{2\alpha}
({\partial}_{\mu}A^{\mu})N - \frac{1}{8\alpha}N^{2} +
\frac{1}{8}(1 - \frac{1}{\alpha})M^{2}\nonumber\\ &+&
\frac{1}{4\alpha}M({\partial}_{++}B_{--}) - \frac{1}{8\alpha}
({\partial}_{++}B_{--})^{2}  + %%@
\frac{1}{8}\biggl(1-\frac{1}{\alpha}\biggl)(\partial_{++}\partial_{--}C)^{2}\nonumber\\ &+& %%@
\frac{1}{4}
\biggl(1+\frac{1}{\alpha}\biggl)(\partial_{++}\partial_{--}C)M-\frac{1}{4\alpha}(\partial_{++}
\partial_{--}C)(\partial_{++}B_{--})\nonumber\\
&-&\frac{i}{2\alpha}({\rho + i\eta}) %%@
\stackrel{\mathrm{\leftrightarrow}}{\partial}_{++}({\bar\rho -i\bar
\eta}) +
\frac{i}{8}{\Sigma}\stackrel{\mathrm{\leftrightarrow}}{\partial}_{++}{\bar\Sigma}\}.  %%@
\label{total} 
\end{eqnarray}

Using  
eq.(\ref{total}), we are ready to write down the propagators for $A^{a}$,
$B^{a}_{--}$, $C^{a}$, $N^{a}$, $M^{a}$, ${\rho}^{a}$, ${\eta}^{a}$, ${\chi}^{a}$ and
${\omega}^{a}$:  
\begin{eqnarray}  
\langle AA \rangle &=& -\frac{2i}{\Box}({\theta}_{\mu\nu} + %%@
2{\alpha}{\omega}_{\mu\nu}),\nonumber\\ 
\langle AN \rangle &=& - \langle N{A} \rangle = %%@
-4i\alpha\frac{\partial_{\mu}}{\Box}\nonumber\\
\langle NN \rangle &=& 8i\alpha \nonumber\\
\langle CC \rangle &=& \frac{32i\alpha}{3\alpha^2+4\alpha+4}\frac{1}{\Box^2},\nonumber\\
\langle BC \rangle &=& -\langle CB \rangle = %%@
-\frac{32i\alpha(3\alpha+2)}{3\alpha^2+4\alpha+4}\frac{\partial_{--}}{\Box^2}\nonumber\\
\langle MC \rangle &=& \langle CM \rangle = %%@
\frac{-32i\alpha(\alpha+1)}{3\alpha^2+4\alpha+4}\frac{1}{\Box}\nonumber\\
\langle BB \rangle &=& -\frac{16i\alpha(\alpha %%@
+2)(3\alpha+10)}{3\alpha^2+4\alpha+4}\frac{{\partial}^{2}_{--}}{{\Box^{2}}},\nonumber\\
\langle MM \rangle &=& \frac{8i\alpha(\alpha+4)}{3\alpha^2+4\alpha+4} \nonumber\\
\langle MB \rangle &=& -\langle BM \rangle =
\frac{-48i\alpha(\alpha+2)}{3\alpha^2+4\alpha+4}\frac{\partial_{--}}{\Box}\nonumber\\
\langle (\rho + i\eta)(\bar\rho -i\bar\eta) \rangle &=&-4\alpha %%@
\frac{\stackrel{\mathrm{\leftrightarrow}}{\partial}_{++}}{\Box}
\nonumber\\
\langle (\rho + i\eta)(\chi +i\omega) \rangle &=&
4\alpha \frac{\stackrel{\mathrm{\leftrightarrow}}{\partial}_{++}}{\Box}\nonumber\\
\langle (\bar\chi - i\bar\omega)(\bar\rho -i\bar\eta) \rangle &=&4\alpha %%@
\frac{\stackrel{\mathrm{\leftrightarrow}}{\partial}_{++}}{\Box}\nonumber\\
\langle (\bar\chi - i\bar\omega)(\chi +i\omega) \rangle &=&
4(\alpha+4) \frac{\stackrel{\mathrm{\leftrightarrow}}{\partial}_{++}}{\Box}.
\label{prop}  
\end{eqnarray} 

One immediately checks that
the extra gauge field, $B_{--}$, does  \underline{not}  decouple from the
matter sector. In the non-Abelian case, the extra gauge potential 
$B_{--}$ also behaves as a second gauge field, exactly as it did in the Abelian case.   
It is very interesting to point out that, in the Abelian case,
$B_{--}$ showed the same behaviour as here : a massless pole of order two \cite{20SYM}; the %%@
difference is that there, it coupled only to $C$ instead of $C$ and $M$, but these two fields %%@
show the same kind of behaviour: they are both compensating fields. Once again, this
ensures us to state that $B_{--}$ behaves as a  physical gauge field: it has
dynamics and couples both to matter and the gauge field $A^{\mu}$. Its  only remaining
peculiarity regards the presence of a single component in the  light-cone
coordinates. 

Let us now turn to the coupling of the two  gauge potentials, $A_{\mu}$ and $B_{--}$, to a
non-linear  $\sigma$-model always keeping a sypersymmetric scenario. It is
our main purpose henceforth to carry out the coupling of a  $(2,0)$
$\sigma$-model to the relaxed gauge superfields of the ref.  \cite{chair}, and
show that the extra vector degrees of freedom do not  decouple from the matter
fields (that is, the target space 
coordinates)\cite{bagger}-\cite{hel}. To perform the coupling of the
$\sigma$-model to the Yang-Mills fields we reason along the same
considerations as i ref.\cite{20SYM} and find out that: \begin{eqnarray} 
{\cal L}_{\xi} &=& {\partial}_{i}[K(\Phi,\tilde\Phi) - {\xi}(\Phi)  -
{\tilde\xi}(\tilde\Phi)]{\nabla}_{--}{\Phi}^{i} + \nonumber\\ &-& 
{\tilde\partial}_{i}[K(\Phi,\tilde\Phi) - {\xi}(\Phi) - 
{\tilde\xi}(\tilde\Phi)]{\nabla}_{--}{\tilde\Phi}^{i},  \label{calL} 
\end{eqnarray}  where ${\xi(\Phi)}$ and ${\bar\xi}(\bar\Phi)$ are a pair of
{\it chiral} and {\it  antichiral} superfields, ${\tilde  \Phi}_{i} \equiv
exp(i{\bf L}_{V.\bar k}){\bar  \Phi}_{i}$ and ${\nabla}_{--}{\Phi}^{i}$
and  ${\nabla}_{--}{\tilde \Phi}^{i}$ are defined in perfect analogy to 
what is done in the case of the bosonic ${\sigma}$-model: 
\begin{eqnarray} 
{\nabla}_{--}{\Phi}_{i} \equiv {\partial}_{--}{\Phi}_{i} - g 
{\Gamma}_{--}^{\alpha} {k}_{\alpha}^{i}(\Phi) 
\end{eqnarray} 
and 
\begin{eqnarray} 
{\nabla}_{--}{\tilde \Phi}_{i} 
\equiv {\partial}_{--}{\tilde \Phi}_{i} - g {\Gamma}_{--}^{\alpha} {\bar 
k}_{\alpha i} (\tilde \Phi). 
\end{eqnarray} 

The interesting point we would like to stress is that the extra 
gauge degrees of freedom accommodated in the component-field 
$B_{--}(x)$ of the superconnection ${\Gamma}_{--}$ behave as a 
genuine gauge field that shares with $A^{\mu}$ the feature of coupling to 
matter and to $\sigma$-model \cite{chair}. This result can be 
explicitly read off from the component-field Lagrangian projected 
out from the superfield Lagrangian ${\cal L}_{\xi}$. We therefore 
conclude that our less constrained $(2,0)$-gauge theory yields a 
pair of gauge potentials that naturally transform under the action 
of a single compact and simple gauge group and may be consistently coupled to
matter fields as well as to the $(2,0)$ non-linear $\sigma$-models by
means of the gauging of their isotropy and isometry groups. 

Relaxing
constraints in the $N=1$- and $N=2-D=3$ supersymmetric algebra of covariant
derivatives may lead to a number of peculiar features of gauged
nonlinear $\sigma$-model\cite{tripa} in the presence of Born-Infeld terms for the
pair of gauge potentials that show up from the relaxation of the superspace constraints\cite{eu}.


\begin{thebibliography}{99} 

\bibitem{20SYM} A. B. Penna-Firme, M. S. G\'oes-Negr\~ao and M. R. Negr\~ao,
{\it ``(2,0)-Super-Yang-Mills Coupled to Non-Linear $\sigma$-Models''}, {\em
Int. J. Mod. Phys.} {\bf A16} (2001) 189.  

\bibitem{hull} C. M. Hull and E.
Witten, {\em Phys. Lett.}{\bf 160B}(1985) 398.   
\bibitem{candelas} P. Candelas, G. Horowitz, A. Strominger and E. Witten, 
{\em Nucl. Phys.}{\bf B258}(1985) 46;\\ C. M. Hull, {\em Nucl. 
Phys.}{\bf B260}(1985) 182 and {\em Nucl. Phys.}{\bf B267}(1986) 
266;\\ A. Sen, {\em Phys. Rev.}{\bf D32}(1985) and {\em Phys. Rev. 
Lett.}{\bf 55}(1985) 1846. 
 
\bibitem{saka} M. Sakamoto, {\em Phys. Lett.}{\bf 
B151}(1985) 115. 
 
\bibitem{gross} D. J. Gross, J. A. Harvey, E. J. Martinec and R. Rohm, {\em 
Phys. Rev. Lett.}{\bf 54}(1985) 502. 
 
\bibitem{porrati} M. Porrati and E. T. Tomboulis, {\em Nucl. 
Phys.}{\bf B 315}(1989) 615;\\ J. Quackenbush, {\em Phys. Lett.}{\bf 
B234}(1990) 285; \\ H. Oguri and C. Vafa, {\em Mod. Phys. Let.}{\bf 
A5} (1990) 1389; {\em Nucl. Phys.}{\bf B361} (1991) 469;{\em Nucl. Phys.}{\bf B367} (1991) 
83;\\ S. J. Gates Jr. and H. Nishino, {\em Mod. Phys. Lett.}{\bf A7} (1992)
2543;  \\ S. J. Gates Jr. S. V. Ketov and H. Nishino, {\em Phys. Lett.}{\bf
B307} (1993) 323;  \\ S. J. Gates Jr. S. V. Ketov and H. Nishino, {\em Phys.
Lett.}{\bf B297} (1992) 99;  {\em Nucl. Phys.}{\bf B393} (1993) 149;\\ M. A.
De Andrade, O. M.  Del Cima and L. P. Colatto, {\em Phys. Lett.}{\bf B370}
(1996) 59.   
\bibitem{chair} N. Chair, J. A. Helay\"el-Neto and A. William Smith, {\em 
Phys. Lett.}{\bf B 233}(1989) 173. 
 
 
\bibitem{dine} M. Dine and N. Seiberg, , {\em Phys. 
Lett.}{\bf 180B}(1986) 364. 
 
\bibitem{bagger} J. Bagger and E. Witten, 
{\em Phys. Lett.}{\bf 118B}(1982) 103. 
 
\bibitem{kar} C. M. Hull, A. 
Karlhede, U. Lindstr\"om and M. Ro\u{c}ek, {\em Nucl. Phys.}{\bf 
B266}(1986) 1;M. Carvalho, J. A. Helay\"el-Neto and M. W. de 
Oliveira, {\em Phys. Rev}{\bf D55} (1997) 7574;\\ M.
Carvalho,  J. A. Helay\"el-Neto and L. C. Q. Vilar, {\em Helv. Phys. Acta}{\bf
71} (1998)  184. 
 
\bibitem{brooks} R. Brooks. F. Muhammad and S. J. Gates 
Jr., {\em Nucl. Phys.}{\bf B268}(1986) 599. 
 
\bibitem{hel} C. A. S. 
Almeida, J. A. Helay\"el-Neto and A. William Smith, {\em Mod. Phys. 
Lett.}{\bf A6}(1991) 1397 and {\em Phys. Lett.}{\bf 279B}(1992) 75. 
 

\bibitem{tripa} P. K. Tripathy, {\em Phys. Rev.}{\bf D59} (1999) 085004.

\bibitem{eu} M. S. G\'oes-Negr\~ao, J. A. Helay\"el-Neto and M. R. Negr\~ao,
{\it work in progress}.
  
\end{thebibliography}
\end{document}